\documentclass[preprint,12pt,authoryear]{elsarticle}
\usepackage{amsmath}
\usepackage{amsfonts}
\usepackage{amssymb}
\usepackage{graphicx}
\usepackage{xcolor}%
\journal{JMPS}
\providecommand{\U}[1]{\protect\rule{.1in}{.1in}}

\begin{document}
\begin{frontmatter}

\title{On mixed-mode fracture mechanics models for {contact area
reduction} under shear load in soft materials}
\author{A.Papangelo$^{(a,b)}$, M. Ciavarella$^{(a,b,\ast)}$}

\address{$^{(a)}$Politecnico di BARI. Center of Excellence in Computational\\Mechanics. Viale Japigia 182, 70126 Bari, Italy\\$^{(b)}$Hamburg University of Technology, Department of Mechanical\\Engineering, Am Schwarzenberg-Campus 1, 21073 Hamburg, Germany\\$^{\ast}$corresponding author: Mciava@poliba.it}

\begin{abstract}
The fundamental problem of friction in the presence of macroscopic adhesion,
as in soft bodies, is receiving interest from many experimentalists. Since the
first fracture mechanics `purely brittle' model of Savkoor and Briggs, models
have been proposed where the mixed mode toughness is interpreted with
phenomenological fitting coefficients introducing weaker coupling between
modes than expected by the "purely brittle\textquotedblright\ model. We
compare here two such previously proposed models and introduce a third one to
show that the transition to sliding is very sensitive to the form of the
mixed-mode model. In particular, after a quadratic decay of the contact area
with load for modest tangential loads, there could be an inflexion point and
an asymptotic limit, or a jump to the Hertzian contact area. We find also that
the unstable points are different under load or displacement control. The idea
that the mixed mode function and parameter should be an interface property may
be erroneous.

\end{abstract}


\end{frontmatter}

\section{Introduction}

{Adhesion and friction are two very discussed topic in tribology,
and the classical Bowden-Tabor view typical for metals was that friction was
largely due to adhesion, the puzzle remaining as to why adhesion could not be
measured in separating the surfaces, while friction clearly was. Today the
same topics receive great attention from the academic community and large
computational models are possible (Vakis et al. (2018), Pastewka \& Robbins
(2014)) which however do not fully answer the very fundamental open questions.
In soft materials where adhesion is more evident, friction is measured also
under zero closure force (Yoshizawa et al. (1993), Homola et al. (1990)), and
hence a fracture mechanics model seems more promising.\ Indeed, }Savkoor
\&\ Briggs (1977) studied the interplay of adhesion and friction for smooth
spheres. They extended the JKR (Johnson et al., 1971) solution for the adhesive
contact of a sphere to the case of tangential loads (see Fig. 1). However,
they found that the contact area reduction was \textit{greatly overestimated}
by this "purely brittle" model in which there is an exact combination of the
modes coming from Linear Elastic Fracture Mechanics theory (Fig. 1 (c)).
However, there are various evidences that in many interface problems, there is
an apparent increase of the toughness of the material under mixed modes, and
there are various phenomenological models to interpret the mode combination
effect. For example, Cao and Evans (1989) experimentally measured the material
toughness for an epoxy-glass bi-material interface, showing that it strongly
increases with the phase angle
\begin{equation}
\psi=\arctan\left(  \frac{K_{II}}{K_{I}}\right)
\end{equation}
being $K_{II}$ and $K_{I}$ respectively the mode II and mode I stress
intensity factors. Physically, this was explained by the fact that,
particularly for low crack opening, several microscopic phenomena affect the
interface toughness, such as friction, plasticity and dislocation emission
(Hutchinson, 1990). The simplest models include one empirical coefficient and
a mode-mixity function $f\left(  \psi\right)  $ (Hutchinson \& Suo, 1992)
where the critical condition for propagation is written as
\begin{equation}
G=G_{Ic}f\left(  \psi\right)  \label{Gpsi}%
\end{equation}
where $G$ is the energy release rate, $G_{Ic}$ is mode I critical factor (or
surface energy, if we assume Griffith's concept) hence $G_{Ic}f\left(
\psi\right)  \ $is the critical energy release rate $G_{c}$ for crack propagation.

Johnson (1996) therefore reconsidered Savkoor \&\ Briggs (1977) model
introducing one such simple mode mixity function, with a quadratic dependence
of the interfacial toughness on $K_{II}/K_{I},$ i.e. writing%
\begin{equation}
f_{a}\left(  \psi\right)  =1+\left(  1-\lambda\right)  \tan\left(
\psi\right)  ^{2}=1+\left(  1-\lambda\right)  \left(  \frac{K_{II}}{K_{I}%
}\right)  ^{2}\qquad\text{(model "a")} \label{fa}%
\end{equation}
\newline and $0\leq\lambda\leq1$ an empirical parameter. The case $\lambda=0$
corresponds to the mode-uncoupling, i.e. to the rather nonphysical condition
for which tangential load doesn't affect at all the adhesion process, and
there is no shrinking of the contact area upon shearing. If $\lambda=1$
instead, $f\left(  \psi\right)  =1$ and this corresponds to the "ideally
brittle" fracture of Savkoor \&\ Briggs (1977) model, where frictional
dissipation is neglected. The empirical constant $\lambda$ permits to tune the
"interaction" between modes and finds much better agreement with the area
decrease with tangential load.

However, (see Fig.3 of Johnson (1996)), we also observe that, except for the
limit case of the "ideally brittle" behavior ($\lambda=1$),  peeling of an
adhesive contact by a tangential force occurs by a monotonically decreasing
contact area with a tendency to show \textit{an asymptotic value of reduction,
and no {jump} instability}. For $\lambda=1$, instead, as Savkoor \&
Briggs (1977) had already noticed, there is a {jump} instability at
a critical tangential force, for which they predicted the Hertz contact radius
was reached. Johnson (1996) disregarded this result because{ as we have
discussed }$\lambda=1$ seems to be a remote limit.%

\begin{center}
\includegraphics[
width=4.9839in
]%
{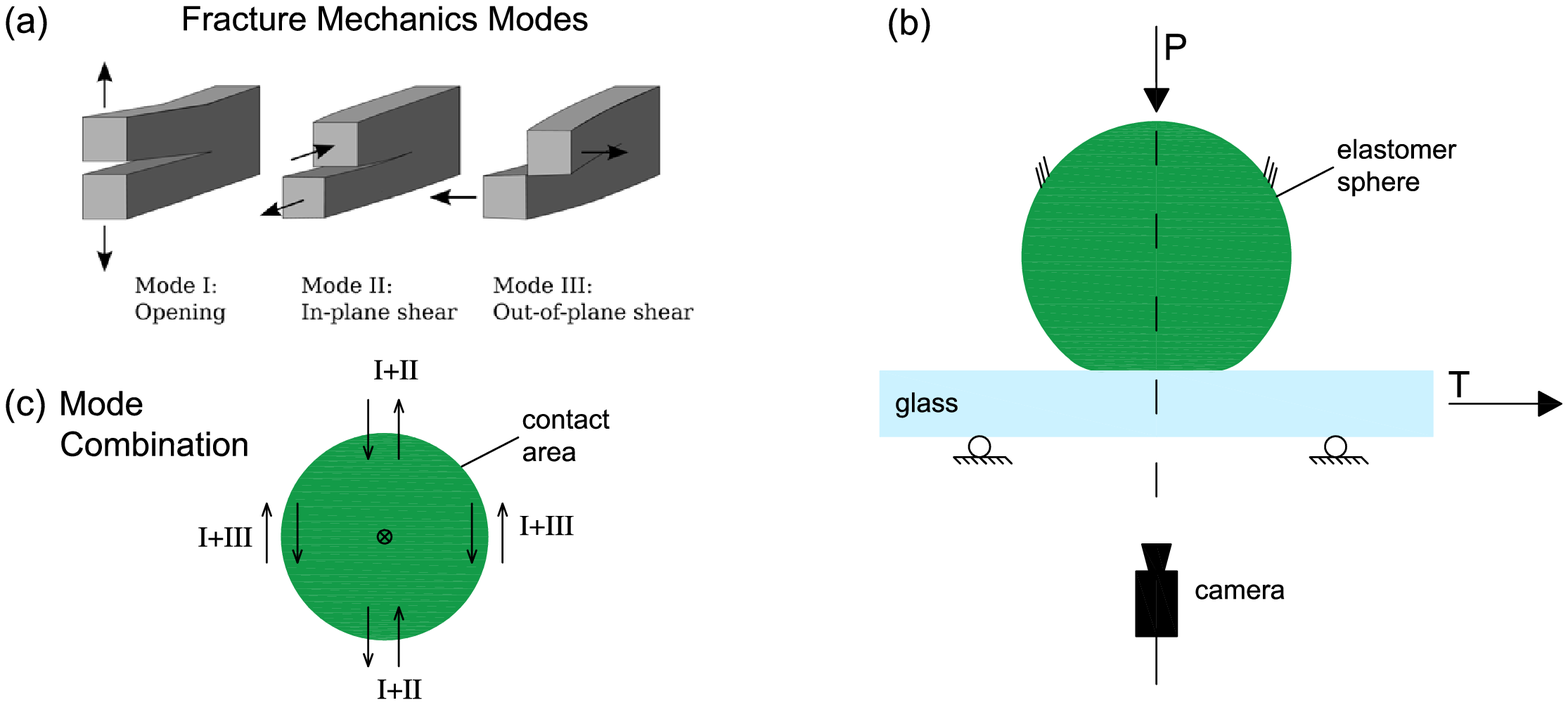}%
\end{center}

\begin{center}
{ Fig. 1 - (a) Fracture mechanics modes: sketch. (b) Schematic
representation of a typical experimental set-up for investigating the
interplay between adhesion and friction in soft materials. Commonly, force
transducers are used to measure normal and tangential loads, while a camera is
used to visualize the macroscopic contact area. (c) Schematic representation
of the combination of modes at the periphery of the contact area in presence
of adhesion and friction}
\end{center}

Then, Johnson (1997) tried to extend this model even further considering a
more "ductile" mode of fracture, introducing a cohesive model in both mode
I\ and mode II, where in particular he modelled the advancement of slip within
the contact area, and assumed another "rather arbitrary" (as he writes) form
for the interaction between modes. This resulted in possible {jump
instabilities} only in the "ductile" regime and not in the JKR one (Johnson
\textit{et al.}, 1971), at least for the cases he considers. However, JKR
equations are known to be valid even in contacts as small as those in an AFM
measurements (Jacobs \& Martini (2017), Ciavarella \& Papangelo (2017)) and
the need of the ductile model (at least, for mode I) appears rather limited.
Therefore, it is worth to investigate if {jump instabilities}
should be predicted by non-cohesive models, perhaps changing the arbitrary
form of the mode mixity function.

Waters \& Guduru (2010) obtained extensive experimental results of the
frictional adhesive contact of a sphere on a plane surface (respectively glass
on PDMS), and found that a mixed mode model like that of Hutchinson and Suo
(1990)
\begin{equation}
f_{b}\left(  \psi\right)  =1+\tan^{2}\left[  \left(  1-\frac{\lambda}%
{2}\right)  \psi\right]  \qquad\text{(model "b")}\label{fb}%
\end{equation}
fitted results quite well\footnote{Notice that for this form of
Waters and Guduru (2010) mode-mixity function, $\lambda$ ranges from $0$ to
$2.$ We changed the definition of $\lambda,$ so that, for $\lambda\ll1,$ all
the three models have the same second order expansion.}, at least until the
contact area remained circular, and beyond this case, they don't attempt
further comparisons. However, beyond this level, cycles of slip instability
and reattachment appeared for compressive normal loads (see Fig. 2), which
should not be confused with Schallamach waves (Schallamach, 1971), but are
rather single slip and reattachment events. These {slip/resticking
instabilities} have been observed by a number of authors, but not all of them,
and hence seem specific to the particular experimental testing apparatus,
method and materials used (see discussion in Waters \& Guduru (2010)), and one
possible source of discrepancy will be given here as depending on the
stiffness of the system, {which could lead to situations closer to "load
control" or to "displacement control".} The amplitude of the tangential force
oscillations is much larger for low normal loads and continuously decrease
with an increase of normal load.%

\begin{center}
\includegraphics[
height=3.2423in,
width=4.9848in
]%
{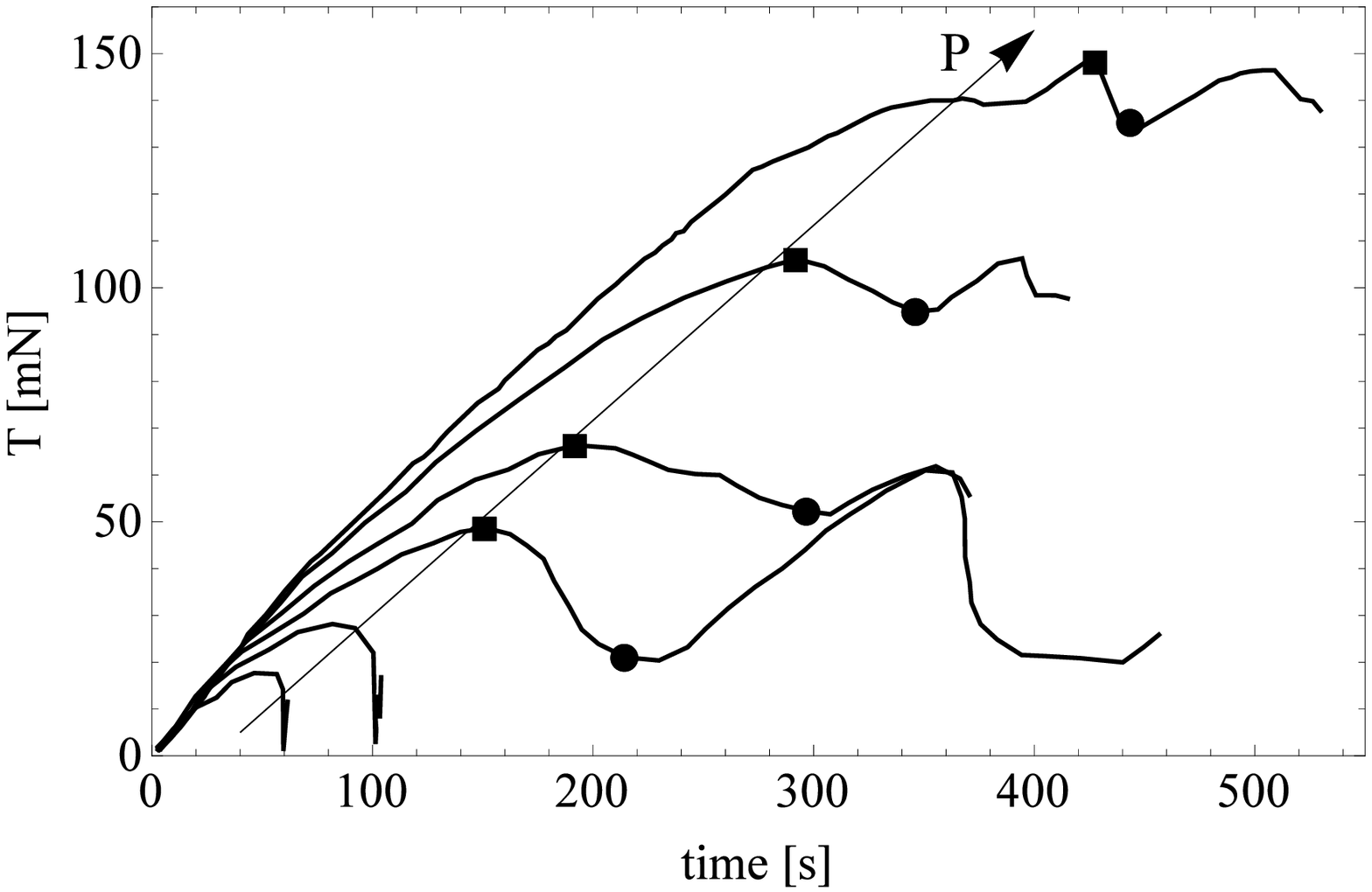}%
\end{center}

\begin{center}
Fig. 2 - Loading curve: tangential load versus time for increasing normal load
$P=\left[  -5.5,-2.5,0.5,3.5,9.5,15\right]  $ mN. Data have been extracted
from Fig. 6 in Waters and Guduru (2010). Squares/circles individuate the
maximum/minimum tangential force in the first cycle of detachment/reattachment.
\end{center}

More recently, Sahli et al. (2018) make measurements on the reduction of the
contact area $A$ upon application of shear force $T$ for a glass-PDMS
interface, suggesting a quadratic decay $A=A_{0}-\alpha_{A}T^{2}$, where
$A_{0}$ is the contact area at $T=0$ and $\alpha_{A}$ is a fitting
coefficient. They also find a scaling law for $\alpha_{A}$ of the form
$\alpha_{A}\sim A_{0}^{-3/2}$ for both macroscopic Hertzian and rough
contacts. Ciavarella (2018) has shown that qualitatively the findings of Sahli
\textit{et al.} (2018) seem justified with the mixed mode fracture mechanics model "a"
but only discussed an asymptotic expansion for low reduction of contact area
and small tangential load. More recently Mergel \textit{et al.} (2018) experiments on a
similar set-up report decay of the contact area which is not really quadratic
for the entire range of observed behavior, as we shall discuss more in details.

It is clear that both the forms proposed by Johnson (1996) and Waters \&
Guduru (2010) (models "a" and "b") imply an unbounded growth of the
interfacial toughness $G_{c}$ under pure mode II, which has to be considered
\textit{unrealistic}. Indeed, Hutchinson \& Suo (1992) warn that some
functional forms "should not be taken literally". \ Hence, in the present
paper, we discuss the implications of adopting another form, also suggested by
Hutchinson \& Suo (1990)%
\begin{equation}
f_{c}\left(  \psi\right)  =\left[  1+\left(  \lambda-1\right)  \sin^{2}\left(
\psi\right)  \right]  ^{-1}\qquad\text{(model "c")}\label{fc}%
\end{equation}
which also corresponds to the very simple model%
\begin{equation}
\frac{1}{2E^{\ast}}\left[  K_{I}^{2}+\lambda K_{II}^{2}\right]  =G_{Ic}%
\label{simple-model}%
\end{equation}
where $E^{\ast}$ is the plain strain elastic modulus $E^{\ast}=\left(
\frac{1-\nu_{1}^{2}}{E_{1}}+\frac{1-\nu_{2}^{2}}{E_{2}}\right)  ^{-1}$ and
$E_{i},$ $\nu_{i}$ are the Young modulus and Poisson's ratio of the material
couple. Fig. 3 shows Johnson's form $f_{a}$\ (dot-dashed black line),
Waters-Guduru's form~$f_{b}$\ (dashed solid line) and the proposed form
$f_{c}$ (red solid line), all with $\lambda=0.15$ which is a realistically low
value to produce low coupling between modes as observed in most experiments.

{For $\lambda\ll1$, the three criteria all have a quadratic form at low
$K_{II}/K_{I}$, as\footnote{Notice that the second order
expansion of $f_{b}\left(  \psi\right)  $ is$\ f_{b}\left(  \psi\right)
\simeq1+\left(  1-\lambda+\lambda^{2}/4\right)  \psi^{2}+o\left(
\lambda\right)  ^{4},$ which for $\lambda\ll1$ reduces to $f_{b}\left(
\psi\right)  \simeq1+\left(  1-\lambda\right)  \psi^{2}.$}}
\begin{equation}
f\left(  \psi\right)  \simeq 1+\left(  1-\lambda\right)  \psi^{2}+O\left(  \psi
^{4}\right)
\end{equation}
and hence they start to differ at $\psi\simeq\pi/4$ or so, which is not necessarily
a high value, as we shall find that the contact area has not changed much
dimension in many cases at this point. Notice that (Ciavarella, 2018)
explained the quadratic decay of the contact area observed in Sahli et al
(2018) with Johnson's model "a", but only with some further simplifying assumptions.

It is therefore under high mode mixity that the various suggestions differ,
and we shall show this has profound implications on the expected behavior of
the contact, particularly on whether there is a smooth or an unstable
transition to macroscopic sliding, where a high phase angle $\psi$ is expected.%

\begin{center}
\includegraphics[
height=3.4225in,
width=4.9848in
]%
{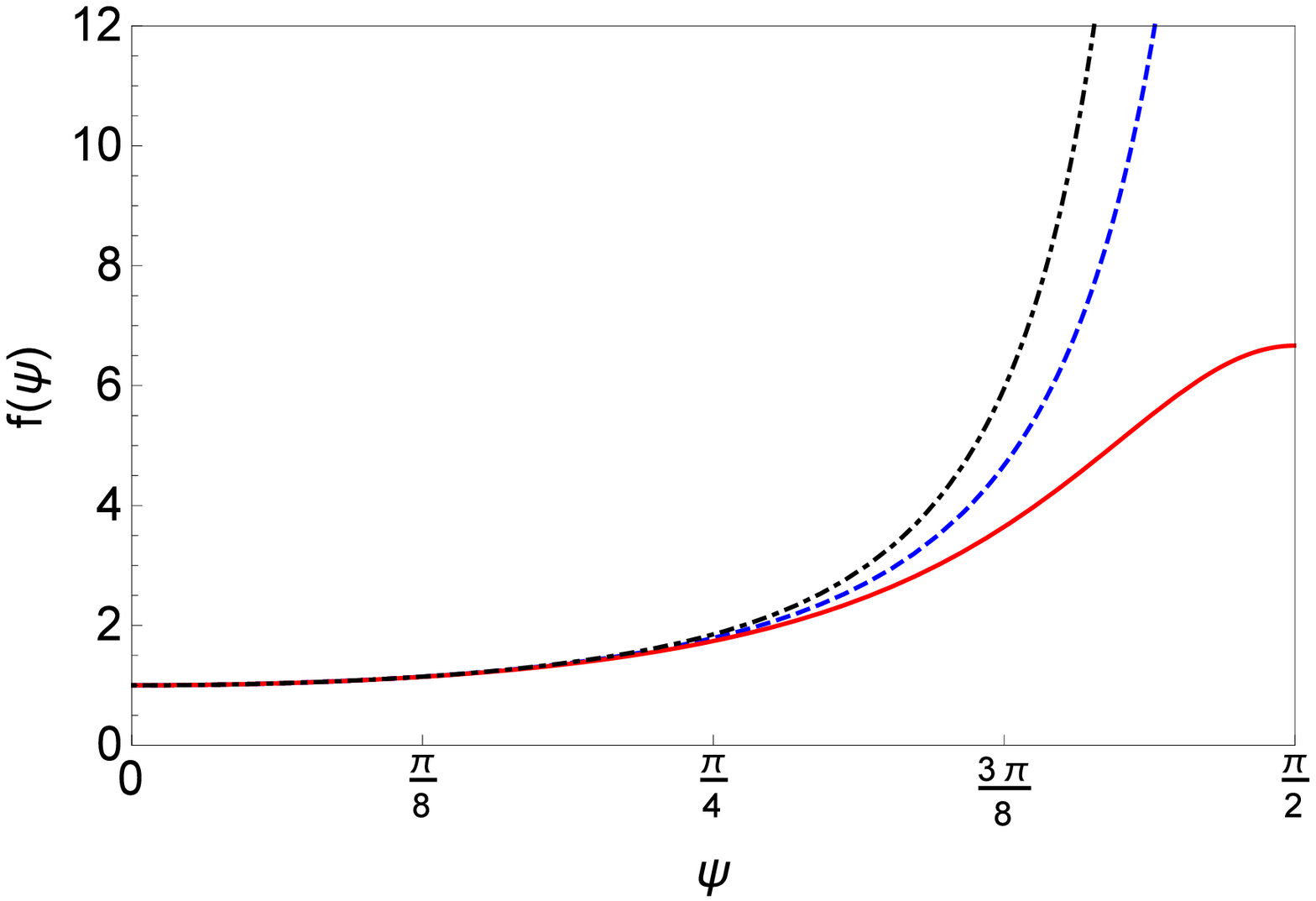}%
\end{center}

\begin{center}
Fig. 3 - Mode mixity growth of critical strain energy release rate $G_{c}$
empirical functions: Johnson's form $f_{a}$\ (dot-dashed black line),
Waters-Guduru's form~$f_{b}$\ (blue dashed line) and the proposed form $f_{c}$
(red solid line), all with $\lambda=0.15$ as a typical value found in
experiments of Waters \& Guduru (2010).
\end{center}

\section{Fracture mechanics model}

\subsection{General calculations}

Consider an Hertzian profile of radius $R$ which indents an halfspace and then
is sheared by a tangential force $T.$ We assume short range adhesion at the
interface, so that the JKR model (Johnson \textit{et al.}, 1971) can be
applied. This is the typical condition of most of the experiments to which we
refer to, including those of Mergel et al.(2018). We assume that upon
shearing, the contact area shrinks, but remains circular with radius $a$
{and no-slip enters within the contact circle}, or more precisely,
that the effect of slip is "included" in the form of the empirical mixed mode
function: as a cohesive model describing slip like Johnson (1997) still
requires an empirical mixed mode function, we find it not a very useful
complication. 

An additional difficulty in this problem arises from the fact that we have an
unusual combination of modes, having all three modes of fracture present in
various degrees along the interface (see Fig. 1(c)). If resistance to mode II and mode III
were equal, we would actually have shrinking starting from the transverse
direction, since mode III has a greater weight in the energy release rate%
\begin{align}
G &  =\frac{1}{2E^{\ast}}\left[  K_{I}^{2}+K_{II\theta}^{2}+\frac{1}{1-\nu
}K_{III\theta}^{2}\right]  \\
K_{II\theta} &  =\frac{T}{2a\sqrt{\pi a}}\cos\left(  \theta\right)  ;\qquad
K_{III\theta}=\frac{T}{2a\sqrt{\pi a}}\sin\left(  \theta\right)  ;
\end{align}
(where $\theta$ is the angle between the radius vector and the direction of
$T$ and $\nu$ is the Poisson ratio) whereas the opposite is found, suggesting
the strength to mode III is much higher. A detailed model requires an
assumption about how to combine mode II\ and mode III, and there is very
little evidence in the literature to do so, and anyway would result
non-axisymmetric. Previous authors suggest to make an average of the mode II
and mode III and this results in a Poisson's ratio coefficient corrective
factor {"$\frac{2-\nu}{2-2\nu}$"}. This small correction is
inconsistent with the assumption of the axisymmetric contact area, which
requires to assume that the toughness in mode III is twice higher than in mode
II (assuming $\nu=1/2$). We find more consistent to make the averaging around
the periphery multiplying the mode III contribution by $1/2$ , so that the
energy release rate is constant along the interface and we obtain an
axisymmetric failure mode, as
\begin{equation}
G=\frac{1}{2E^{\ast}}\left[  K_{I}^{2}+K_{II}^{2}\right]  \label{1}%
\end{equation}
For mode I, the stress intensity factor is given by%
\begin{equation}
K_{I}=\frac{P_{a}}{2a\sqrt{\pi a}}=\frac{P_{H}-P}{2a\sqrt{\pi a}}\label{Ki}%
\end{equation}
where $a$ is the contact radius and we have split the total load
$P=P_{H}-P_{a}$ into two contributions: a compressive Hertzian load
$P_{H}=\frac{4E^{\ast}a^{3}}{3R}$ and a Boussinesq flat punch solution with
total load $P_{a}$ which is responsible of the contact edge singularity.

In the absence of tangential force, equilibrium dictates $G=G_{Ic}$, and the
standard JKR equation (\ref{1}) gives the contact radius $a$ for a given
normal force $P$
\begin{equation}
P=\frac{4E^{\ast}a^{3}}{3R}-\sqrt{8\pi E^{\ast}a^{3}G_{Ic}}\label{JKR}%
\end{equation}
With applied tangential force $T$, we assume the surfaces do not
slip\footnote{We have shown elsewhere (Papangelo et al. 2015) that a Coulomb
frictional model involving a slip dependent friction coefficient involving
microslip can lead to LEFM singular field in the limit of a small process
zone. For soft materials, generally a constant shear strength is assumed in
the slip zones, which corresponds to the singular mode strictly only in this
limit. }, {which implies a singular field of shear tractions at the
interface}%

\begin{equation}
q\left(  r\right)  =\frac{q_{0}}{\sqrt{1-\left(  \frac{r}{a}\right)  ^{2}}}
\label{qr}%
\end{equation}
{where $r$ is the radial coordinate and $q_{0}=\frac{T}{2\pi a^{2}%
}$.} {The distribution (\ref{qr}) gives rise to a uniform
tangential displacement "$u$" within the loaded circle equal to $\left(
\nu=1/2,\text{ Johnson (1985)}\right)  $
\begin{equation}
u=\frac{3\pi}{2}\frac{q_{0}a}{E^{\ast}}=\frac{3}{4}\frac{T}{aE^{\ast}}
\label{u}%
\end{equation}
} The mode II stress intensity factor $K_{II}$ along the shearing direction is
given by
\begin{equation}
K_{II}=\frac{T}{2a\sqrt{\pi a}} \label{2}%
\end{equation}
Using (\ref{Gpsi}) and (\ref{1}), the critical condition is written as%

\begin{equation}
\frac{1}{2E^{\ast}}\left[  K_{I}^{2}+K_{II}^{2}\right]  =G_{Ic}f\left(
\psi\right)  \label{Gf}%
\end{equation}
which using (\ref{Ki}) and (\ref{2}) leads to%

\begin{equation}
P=\frac{4E^{\ast}a^{3}}{3R}-\sqrt{8\pi E^{\ast}G_{Ic}a^{3}f\left(
\psi\right)  -T^{2}} \label{Eq}%
\end{equation}
\qquad Equation (\ref{Eq}) is the fundamental equation that governs the
contact area reduction while the shear load is increased, for any mode mixity
function $f\left(  \psi\right)  .$

In what follows we will adopt a dimensionless notation, as introduced by
Maugis (2000)%

\begin{align}
\xi &  =\left(  \frac{E^{\ast}R}{G_{Ic}}\right)  ^{1/3};\text{\quad\quad
}\widetilde{a}=\frac{\xi a}{R};\text{\quad\quad}\widetilde{u}=\frac{u\xi^{2}%
}{R};\\
\widetilde{T}  &  =\frac{T}{RG_{Ic}};\text{\quad\quad}\widetilde{P}=\frac
{P}{RG_{Ic}};\text{\quad\quad}\widetilde{q}_{0}=\frac{q_{0}\xi}{E^{\ast}}%
\end{align}
thus the JKR load-contact radius equation becomes simply%

\begin{equation}
\widetilde{P}=\frac{4}{3}\widetilde{a}^{3}-\sqrt{8\pi\widetilde{a}^{3}}
\label{JKRadim}%
\end{equation}
and the governing equation (\ref{Eq}) reduces to%

\begin{equation}
\widetilde{P}=\frac{4}{3}\widetilde{a}^{3}-\sqrt{8\pi\widetilde{a}^{3}f\left(
\psi\right)  -\widetilde{T}^{2}} \label{eqdless}%
\end{equation}
Using eq. (\ref{fa},\ref{fb},\ref{fc}) into (\ref{eqdless}), the governing
equations for the three models considered are obtained, in particular, as

\begin{itemize}
\item Johnson's model (model "a")%
\begin{equation}
\widetilde{P}=\frac{4}{3}\widetilde{a}^{3}-\sqrt{8\pi\widetilde{a}^{3}\left[
1+\left(  1-\lambda\right)  \frac{\widetilde{T}^{2}}{\left(  \frac{4}%
{3}\widetilde{a}^{3}-\widetilde{P}\right)  ^{2}}\right]  -\widetilde{T}^{2}}
\label{Jomodel}%
\end{equation}

\item Waters and Guduru's model (model "b")%
\begin{equation}
\widetilde{P}=\frac{4}{3}\widetilde{a}^{3}-\sqrt{8\pi\widetilde{a}^{3}\left\{
1+\tan^{2}\left[  \left(  1-\frac{\lambda}{2}\right)  \arctan\left(
\frac{\widetilde{T}}{\frac{4}{3}\widetilde{a}^{3}-\widetilde{P}}\right)
\right]  \right\}  -\widetilde{T}^{2}} \label{WGmodel}%
\end{equation}

\item proposed model (model "c"), which turns out to have also the simplest
form%
\begin{equation}
\widetilde{P}=\frac{4}{3}\widetilde{a}^{3}-\sqrt{8\pi\widetilde{a}^{3}%
-\lambda\widetilde{T}^{2}} \label{prop}%
\end{equation}

\end{itemize}

We first plot\ results in the form of contact radius vs normal load curves for
various tangential loads $\widetilde{T}=\left[  10,20,30,40\right]  $ for Fig.
4 (a,c), $\widetilde{T}=\left[  10,40,70,100\right]  $ for Fig. 4 (b) and
$\lambda=0.15$ as a representative value permitting weak coupling between
modes --- the single value obviously fits the same data only at low
$K_{II}/K_{I}$ values, see Fig. 3. All curves in Fig. 4 (a-b-c correspond to
model a,b,c respectively) lie in between the JKR (dot-dashed) and Hertz
(dashed)\ known limits. However, the detailed form of the mixed mode functions
varies significantly the response, particularly in the way the solution
remains similar to JKR, with just a smooth but small reduction of the contact
area as in models a and b, or tends to Hertz with an unstable point. If the
unstable point is in the tensile region, obviously this predicts a jump off
contact, whereas in cases when this occurs in the area of compressive loads,
we would expect a jump to Hertz contact. We notice in particular the following features:

\begin{itemize}
\item Johnson\ (1996) model "a" (Fig. 4 (a)) shows virtually only one possible
{detachment instability at negative loads,} very close to the
{JKR pull-off, and the curve contact radius vs normal load,
increasing the tangential load, rapidly converges to a limit curve.} There is
therefore no {jump instability} in the compressive
region\footnote{Except for the limit case of $\lambda=1$};

\item Waters-Guduru model "b" (Fig. 4 (b)) shows a {jump
instability, which persists to some extent also for compressive loads and then
becomes a smooth transition into Hertz regime;}

\item The "proposed" model \textquotedblleft c\textquotedblright(Fig. 4 (c))
shows a more marked {jump instability which is present for
\textit{all} values of normal load. For tensile loads, at the jump instability
point, detachment occurs (jump off contact). For compressive loads, at the
jump instability point, the contact area drops in size to the Hertzian value;}
\end{itemize}

\begin{center}%
\begin{center}
\includegraphics[
height=3.5447in,
width=4.9848in
]%
{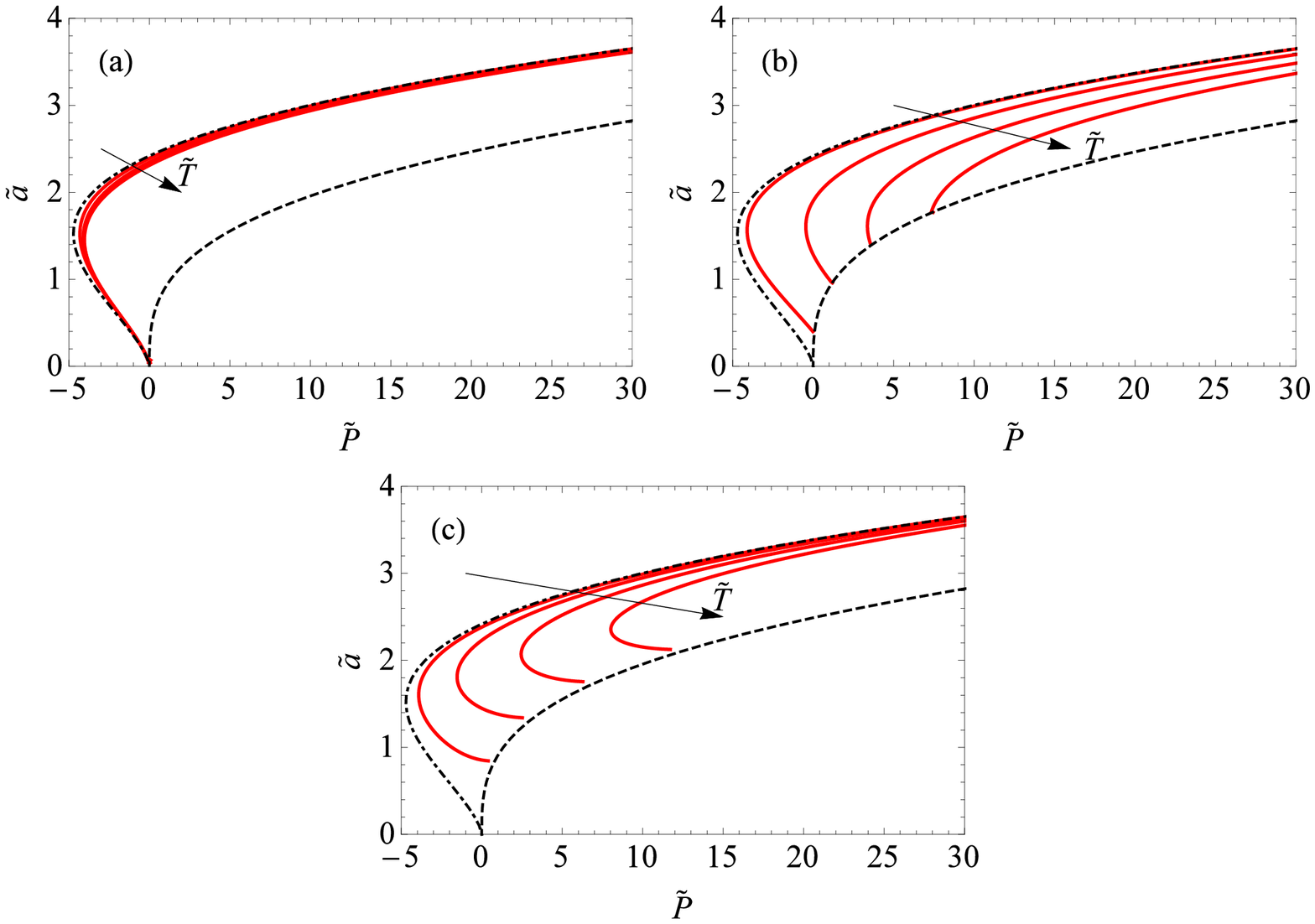}%
\end{center}

Fig. 4 - Contact radius as a function of normal load for $\lambda=0.15$ and
$\widetilde{T}=\left[  10,20,30,40\right]  $ for (a,c) while $\widetilde
{T}=\left[  10,40,70,100\right]  $ for (b). All curves develop between JKR
(dot-dashed) and Hertz (dashed)\ curves for (a) Johnson (1996) model
(b)\ Waters-Guduru model (c) proposed model.
\end{center}

{In Fig. 5 the possible qualitative behaviors are summarized.
Starting at zero tangential load, the solution corresponds to that of JKR
(Johnson \textit{et al.}, 1971). Upon shearing the contact area shrinks: 

\begin{enumerate}
\item continuously up to a non Hertzian solution (Fig. 5, label (1));

\item continuously up to the Hertzian solution (Fig. 5, label (2));

\item continuously up to critical value and then it drops suddenly up to the
Hertzian value (Fig. 5, label (3));
\end{enumerate}

Notice that some authors (Mergel \textit{et al.,} 2018) have suggested that
full-sliding takes place upon reaching a critical interfacial shear strength
$\tau_{0}$ (a property of the interface) (Fig. 5, label (4)), thus we keep the
conservative view to consider two competing mechanisms: fracture mechanics
ones described in (1-2-3) and the shear-strength criterion $T_{full}=$
$\tau_{0}A$ (Fig. 5, label (4)). }

\begin{center}%
\begin{center}
\includegraphics[
height=3.2594in,
width=4.9848in
]%
{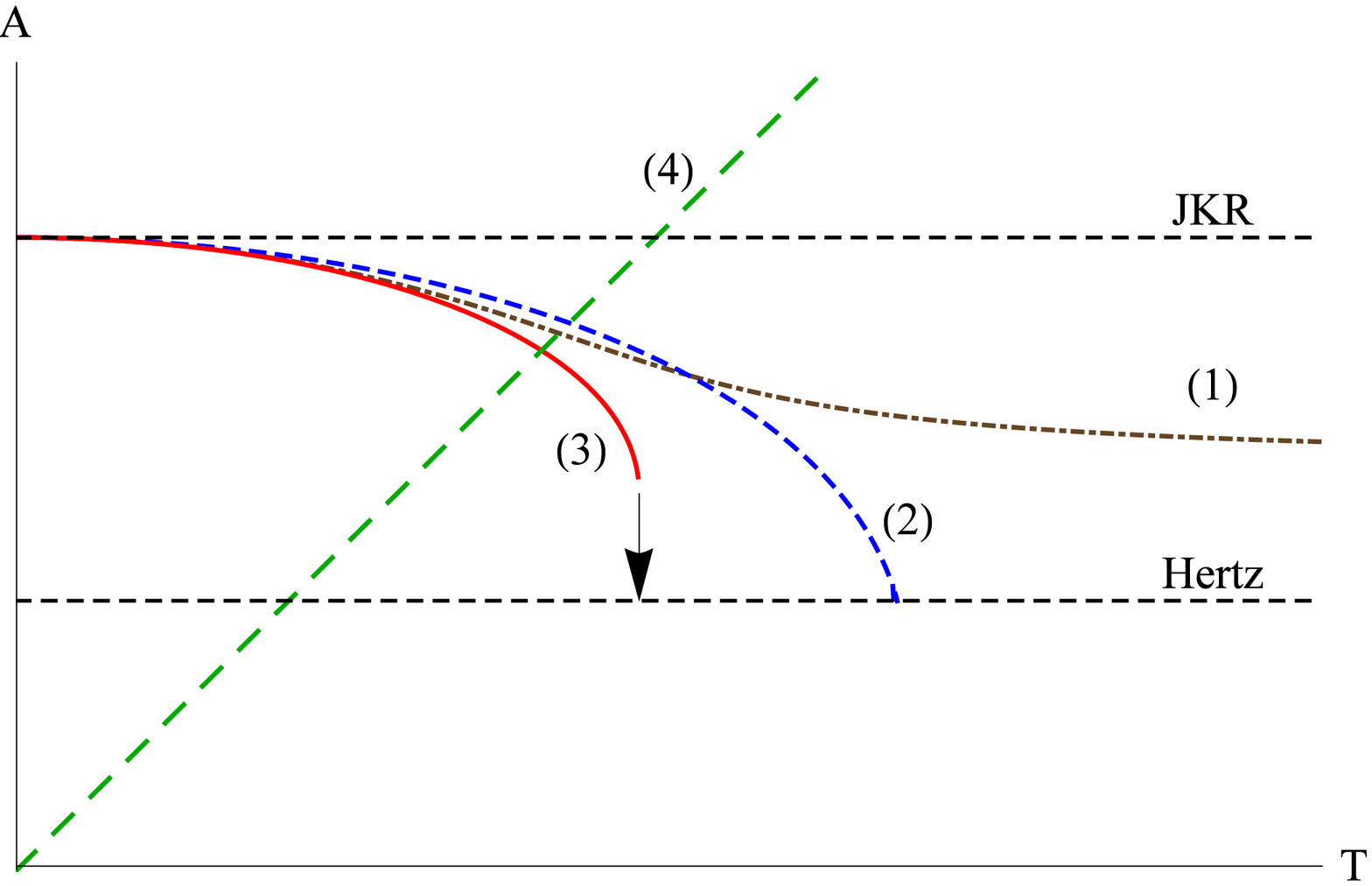}%
\end{center}

{Fig. 5 - Contact area vs tangential load. The possible
qualitative behaviors are plotted and labeled with (1-3). In particular, upon
shearing the contact area may shrink continuously up to a non Hertzian
solution (1), continuously up the the Hertzian solution (2), continuously up
to a critical point and then with a jump instability up to the Hertzian
solution (3). Some authors have proposed that full sliding happens when a
critical shearing traction is reached at the interface $\tau_{0}=T_{full}/A$ (4).}
\end{center}

\subsection{Loading curves}

{For given normal load $\widetilde{P}$ we look for the equilibrium
relation between the tangential load $\widetilde{T}$ and the uniform
tangential displacement $\widetilde{u}.$ Substituting eq. (\ref{u}) into
(\ref{prop}), after some algebra, the relation $\widetilde{u}\left(
\widetilde{T}\right)  $ is obtained analytically (for model "c")}%

\begin{equation}
\widetilde{u}=\frac{3^{2/3}}{2}\left(  \frac{3\pi+\widetilde{P}\pm\sqrt
{9\pi^{2}-\lambda\widetilde{T}^{2}+6\pi\widetilde{P}}}{2\lambda\widetilde
{T}^{2}+2\widetilde{P}^{2}}\right)  ^{1/3}\widetilde{T} \label{loadcurve}%
\end{equation}
where\ we left a "$\pm$" sign, which indicates two different branches.

\begin{center}
\includegraphics[
height=9.3639in,
width=4.9848in
]{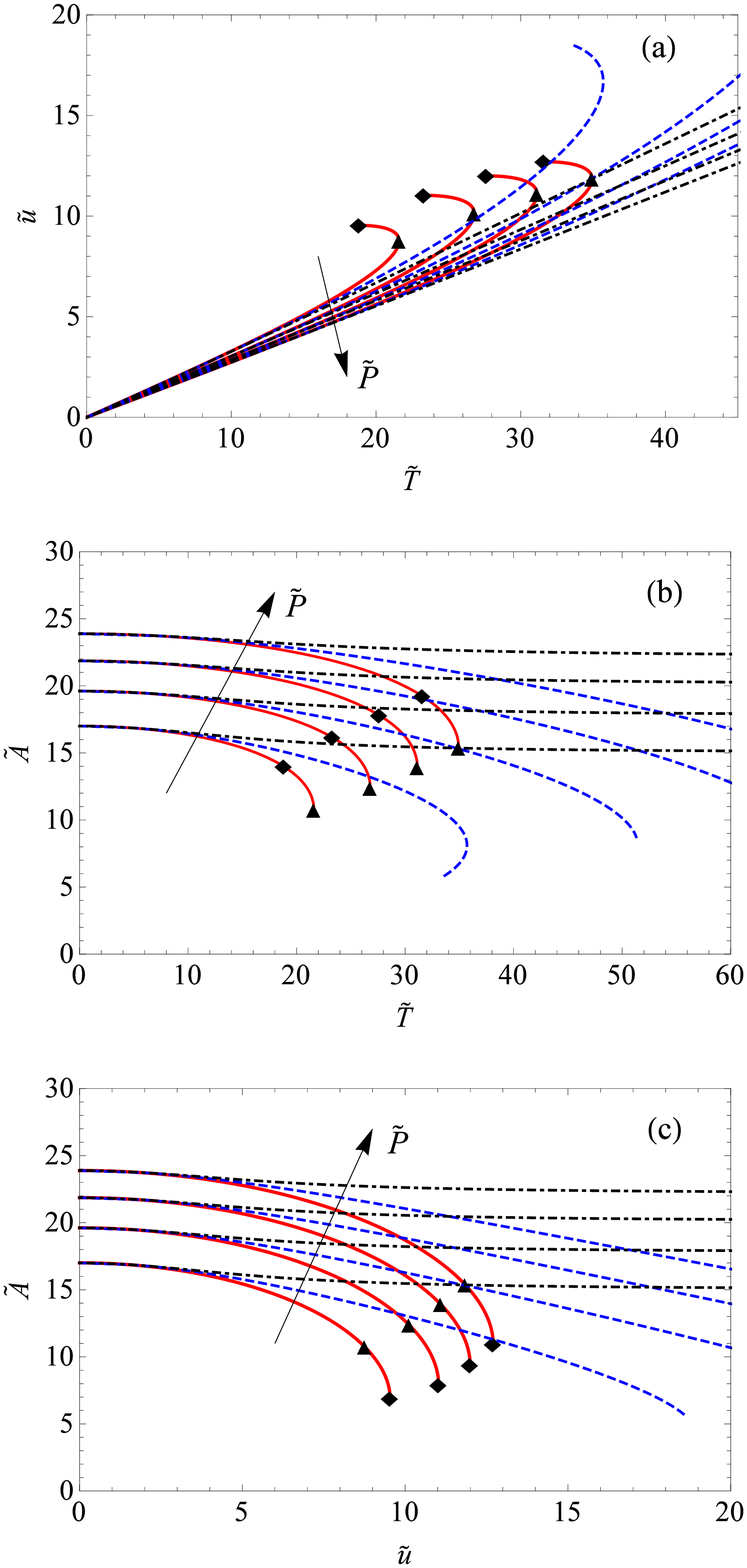}
\end{center}


\begin{center}
{Fig. 6 - (a) Tangential displacement vs tangential load, (b) contact area vs
tangential load and (c) contact area vs tangential displacement for
$\widetilde{P}=\left[  -1,1,3,5\right]  $ and $\lambda=0.15.$ In all panels
black dot-dashed line for model "a" Johnson (1996), blue dashed line for model
"b" Waters-Guduru (2010) and red solid line for the proposed model "c". For
the proposed model "c", the instability points under load (triangles) and
displacement (diamonds) control are indicated.}
\end{center}

{In Fig. 6 the loading curves (panel (a) $\widetilde{u}$ vs $\widetilde{T},$
panel (b) $\widetilde{A}$ vs $\widetilde{T},$ panel (c) $\widetilde{A}$ vs
$\widetilde{u}$) are plotted using $\widetilde{P}=\left[  -1,1,3,5\right]  $
and $\lambda=0.15$ for the three models considered: Johnson model "a" (black
dot-dashed line), Waters-Guduru model "b" (blue dashed line) and proposed
model "c" (red solid line). Notice that only the physically meaningful
branches have been drawn. In Fig. 6 (a) the equilibrium relation
$\widetilde{u}\left(  \widetilde{T}\right)  $ is plotted. While for the model
"c" we have derived the relation $\widetilde{u}\left(  \widetilde{T}\right)  $
analytically (\ref{loadcurve}), for the two models "a" and "b" the curves have
been obtained numerically using eq. (\ref{u}) and (\ref{prop}). Immediately
one notice that the three models behave very differently. In particular for
the Johnson model $\widetilde{u}\left(  \widetilde{T}\right)  $ may be well
approximated by a linear relation, for Waters-Guduru model the curve
$\widetilde{u}\left(  \widetilde{T}\right)  $ starts to bend nonlinearly, and
the proposed model "c" shows two marked instability points, one under load
(triangle symbols), and the other under displacement  control (diamond symbols).}

Fig. 6 (b-c) plots the contact area $\widetilde{A}=\pi\widetilde{a}^{2}$ vs
tangential load $\widetilde{T}$ and displacement $\widetilde{u}$.\ Fig. 6
(b-c) shows that Johnson (1996) model "a" is almost insensitive to the
tangential load (displacement) with this value of $\lambda$: this is because
the toughness grows already very significantly (see Fig. 3). This explains why
we did find $\widetilde{u}$ to have almost a linear dependence on
$\widetilde{T}.$ At the other extreme, with the proposed model "c" , the area
decays in a quasi-elliptical fashion and shows the marked {jump
instability} we have already discussed at a critical value of the tangential
load (Fig. 6 (b-c)). An intermediate behavior is shown by the Waters-Guduru
model.  

Solving in particular eq. (\ref{prop}) for $\widetilde{a},$ the contact area
$\widetilde{A}$ is found as%

\begin{equation}
\widetilde{A}=\pi\left(  \frac{3}{4}\right)  ^{2/3}\left[  3\pi+\widetilde
{P}\pm\sqrt{9\pi^{2}+6\pi\widetilde{P}-\lambda\widetilde{T}^{2}}\right]
^{2/3} \label{Atilde}%
\end{equation}
where the "$\pm$" indicates two possible branches\footnote{In Fig. 6 we drew
only the physically meaningful branches.}. Sahli \textit{et al.} (2018)
suggested a quadratic decay for the contact area with the tangential load
$\ \widetilde{A}=\widetilde{A}_{0}-\widetilde{\alpha}_{A}\widetilde{T}^{2}$
(where $\widetilde{A}_{0}$ is the dimensionless contact area for null
tangential load and $\widetilde{\alpha}_{A}=\alpha_{A}\xi^{2}G_{Ic}^{2}$ is a
fitting parameter) and in view of the fact that all models are similar at low
tangential load, we could expand in series eq. (\ref{Atilde}) up to the second
order to find an analytical expression for $\widetilde{\alpha}_{A}.$ This gives%

\begin{equation}
\widetilde{\alpha}_{A}=\frac{\lambda}{6\times3^{2/3}\left(  2\pi\right)
^{1/3}\sqrt{1+\frac{2\widetilde{P}}{3\pi}}\left(  1+\frac{\widetilde{P}}{3\pi
}+\sqrt{1+\frac{2\widetilde{P}}{3\pi}}\right)  ^{1/3}} \label{alpha}%
\end{equation}
Sahli et al. (2018) showed a scaling law for $\alpha_{A}$ with respect to the
contact area for null tangential load $A_{0},$ over four orders of magnitude.
Using $\widetilde{P}=\frac{4}{3}\left(  \frac{\widetilde{A}_{0}}{\pi}\right)
^{3/2}-\sqrt{8\pi}\left(  \frac{\widetilde{A}_{0}}{\pi}\right)  ^{3/4},$ eq.
(\ref{alpha}) is rewritten as%

\begin{equation}
\widetilde{\alpha}_{A}=\frac{\pi^{7/4}\lambda}{2\times2^{1/3}g\left(
\widetilde{A}_{0}\right)  \left[  4\widetilde{A}_{0}^{3/2}+3\pi^{5/4}\left(
3\pi^{5/4}-2\sqrt{2}\widetilde{A}_{0}^{3/4}+g\left(  \widetilde{A}_{0}\right)
\right)  \right]  ^{1/3}}%
\end{equation}
where $g\left(  \widetilde{A}_{0}\right)  =\sqrt{8\widetilde{A}_{0}%
^{3/2}-12\sqrt{2}\pi^{5/4}\widetilde{A}_{0}^{3/4}+9\pi^{5/2}}$. In the limit
of high $\widetilde{A}_{0}$%

\begin{equation}
\widetilde{\alpha}_{A}\simeq\frac{\pi^{7/4}\lambda}{8\sqrt{2}}\widetilde
{A}_{0}^{-5/4}%
\end{equation}
which predicts a power law decay with exponent $-5/4$ which agrees with
Ciavarella (2018) and is close to $-3/2$ found by Sahli et al. (2018) in their experiments.

{We can also obtain exact results for the {jump
instability points} in the proposed model "c". Under displacement control, at
the {jump instability} point (diamonds in Fig. 6), the derivative
of the tangential displacement with respect to the tangential load
$\frac{d\widetilde{u}}{d\widetilde{T}}$ vanishes, thus the critical tangential
load $\widetilde{T}_{u}$, tangential displacement $\widetilde{u}_{u}$ and
contact area $\widetilde{A}_{u}$ at the {jump instability} points
under displacement control are obtained}%

\begin{align}
\widetilde{T}_{u}  &  =\frac{1}{2}\sqrt{\frac{3}{2}\frac{9\pi^{2}%
+6\pi\widetilde{P}-3\widetilde{P}^{2}+\sqrt{3\left(  3\pi+\widetilde
{P}\right)  ^{2}\left(  3\pi^{2}+2\pi\widetilde{P}+3\widetilde{P}^{2}\right)
}}{\lambda}}\label{Tcru}\\
\widetilde{u}_{u}  &  =\left(  \frac{3}{4}\right)  ^{2/3}\frac{\widetilde
{T}_{u}}{\left(  3\pi+\widetilde{P}+\sqrt{9\pi^{2}+6\pi\widetilde{P}%
-\lambda\widetilde{T}_{u}^{2}}\right)  ^{1/3}}\label{ucru}\\
\widetilde{A}_{u}  &  =\pi\left(  \frac{3}{4}\right)  ^{2/3}\left[
3\pi+\widetilde{P}\pm\sqrt{9\pi^{2}+6\pi\widetilde{P}-\lambda\widetilde{T}%
_{u}^{2}}\right]  ^{2/3} \label{Acru}%
\end{align}

Under load control, at the {jump instability} point (triangles in
Fig. 6), the derivative of the contact area with respect to the tangential
load $\frac{d\widetilde{A}}{d\widetilde{T}}$ is singular, thus the critical
tangential load $\widetilde{T}_{T}$, tangential displacement $\widetilde
{u}_{T}$ and contact area $\widetilde{A}_{T}$ at the {jump
instability} points under load control are obtained%

\begin{align}
\widetilde{T}_{T} &  =\sqrt{\frac{3\pi\left(  3\pi+2\widetilde{P}\right)
}{\lambda}}\label{Tcr}\\
\widetilde{u}_{T} &  =\left[  \frac{\left(  3/4\right)  ^{2}}{3\pi
+\widetilde{P}}\right]  ^{1/3}\sqrt{\frac{3\pi\left(  3\pi+2\widetilde
{P}\right)  }{\lambda}}\\
\widetilde{A}_{T} &  =\pi\left(  \frac{3}{4}\right)  ^{2/3}\left(  \frac{3}%
{2}\pi+\lambda\frac{\widetilde{T}_{T}^{2}}{6\pi}\right)  ^{2/3}\label{Acr}%
\end{align}
From eq. (\ref{Acr}), for $\widetilde{T}_{T}\gg3\pi/\sqrt{\lambda}$, the
critical tangential load scales as $\widetilde{T}_{T}\rightarrow\left(
\frac{\widetilde{A}_{T}}{0.37\lambda^{2/3}}\right)  ^{3/4}$ which is close to
the linear increase of the alternative criterion of full sliding proposed, for
example, by Sahli \textit{et al.} (2018) and Mergel \textit{et al.} (2018) $\widetilde{T}%
_{full}=\widetilde{\tau}_{0}\widetilde{A}$, where $\widetilde{\tau}_{0}%
={{\tau}_{0}\xi}/{E^{\ast}}$ is an interfacial property. Whether the
{jump instability} really occurs or not may not be a trivial
question to answer experimentally, and obviously may depend on the particular
experimental conditions and specific material. As we have also obtained that
the{ jump instabilities depend on the stiffness of the system (with load
control being valid in the limit of soft loading setup and displacement
control for very stiff loading setup), this is also a possible explanation of
the disagreement over the occurrence of the cycles of slip and reattachment
cycles reported by some but not all authors (see the discussion in the paper
of Waters-Guduru (2010)).}

{Also, from Fig. 4 we conclude that the jump
instability will appear at most only at low normal loads, and then it should
disappear for two reasons: one is the absence of the jump instability point in
many mixed mode models, and the other is that, for high normal loads, Hertz
contact area will be close anyway to the area with adhesion (the leading term
in eq. (\ref{JKRadim}) is the adhesionless one \textquotedblleft$\widetilde
{a}^{3}$\textquotedblright). This is in agreement with the observation of
Waters-Guduru (2010), who observed in their Fig. 6 that the jump instability
reduces increasing the normal load (see Fig. 1). }

\section{Detailed comparison with experimental results}

We discuss some comparisons of the models predictions with the experimental
results published by Mergel \textit{et al.} (2018, their Fig. 5 (c)). From
Mergel \textit{et al.} (2018) the following parameters can be extracted:
$G_{Ic}=27$ mJ/m$^{2}$, $R=9.42$ mm, $E^{\ast}=2.133$ MPa. To assess the mode
mixity function, we use the fracture mechanics model rewritten in the form%

\begin{equation}
\frac{G_{c}}{G_{Ic}}=f\left(  \psi\right)  =\frac{\left(  \frac{4E^{\ast}%
a^{3}}{3R}-P\right)  ^{2}+T^{2}}{8\pi E^{\ast}a^{3}G_{Ic}}%
\end{equation}
so that the ratio $G_{c}/G_{Ic}$ as a function of $\psi$ can be obtained
directly from the experiments (Fig. 7, gray symbols) and compared with the
proposed model (solid red line), Waters \& Guduru (dashed blue line) and
Johnson (1966) (dot-dashed black line). The extremely low value of
$\lambda=0.0023$ (coefficient of determination $R^{2}=0.9999\ $) has been
obtained using a least-square best fit procedure of the proposed model
$f_{c}\left(  \psi\right)  $, using the data in the form $\log\left(
\frac{G_{c}}{G_{Ic}}\right)  $ vs $\psi$. We assume the same $\lambda$ holds
for the other two criteria since they have the same form to the second order.

\begin{center}%
\begin{center}
\includegraphics[
height=3.2936in,
width=4.9848in
]%
{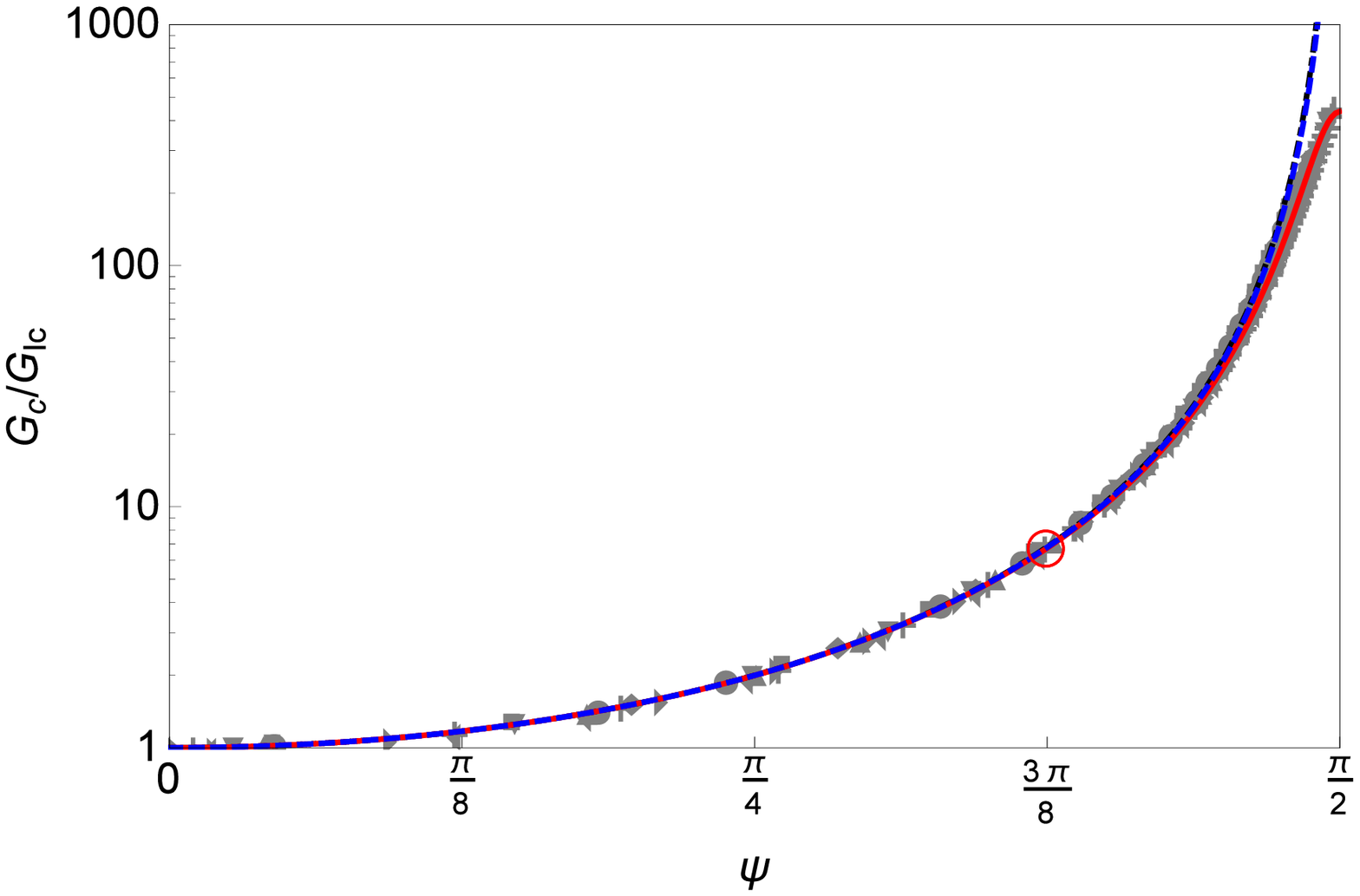}%
\end{center}

Fig. 7 - $G_{c}/G_{Ic}$ is plotted for the experimental data (gray symbols) in
Mergel \textit{et al.} (2018, their Fig. 5 (c)). The dot-dashed, dashed, solid
lines stand respectively for the Johnson (1966) model "a", Waters \& Guduru
model "b", proposed model "c". {Notice that the models
\textquotedblleft a\textquotedblright\ and \textquotedblleft
b\textquotedblright, respectively dot-dashed and dashed lines, are superposed
and almost indistinguishable.} The values of $\lambda=0.0023$ and the
coefficient of determination $R^{2}=0.9999\ $has been obtained using a
least-square best fit procedure of the proposed model $f_{c}\left(
\psi\right)  $ using the data in the form $\log\left(  \frac{G_{c}}{G_{Ic}%
}\right)  $ vs $\psi$. We assume the same $\lambda$ holds for the other two
criteria since they have the same form to the second order.
\end{center}

The best results have been obtained with the suggested model $f_{c}\left(
\psi\right)  .$ However, this is not so evident in the energy release rate
fit, as it is from the contact area predictions in Fig 8. With the fitted
$\lambda$ we directly compare the experimental area vs tangential load curves
with the analytical predictions respectively for (a) Johnson (1996) model
(b)\ Waters-Guduru model and (c) proposed model. It is clearly shown that the
suggested mode-mixity function $f_{c}\left(  \psi\right)  $ fits the
experimental data of Mergel \textit{et al.} (2018) much better than the
Johnson (1966) and Waters \& Guduru (2010) model. {Notice that the
poor fit of the latter models "a" and "b" is not due to our choice of a single
$\lambda$ - we have attempted to tune $\lambda$ to best fit the data in the
entire range but this resulted always in a poor fitting, with low coefficient
of determination.} Instead, with the model "c", the decay is better fitted in
a much wider range, some discrepancies arising only very close to the
{jump instability} point. In all the panels (a, b, c) we have
highlighted with a red circle the point which corresponds to $\psi=\frac{3\pi
}{8}$ and the same red circle is shown in Fig. 7, to show that  most of the
experimental points lie in the region of high $\psi,$ where the three models
presented differ considerably.

Fig. 8 (c) shows a dot-dashed line for the full sliding criterion
$T_{full}=\tau_{0}A$, with $\tau_{0}=0.43$ MPa (the value reported by Mergel
\textit{et al. }(2018)), while the dashed line stands for eq. (\ref{Acr}). It
is clear that the two criteria are almost indistinguishable in Mergel
\textit{et al.} (2018) experimental conditions, so that it is not safe to make
a statement as to which criterion is met. Mergel \textit{et al.} (2018)
conclude for the strength based criterion, but we find this not so obvious.

\begin{center}%
\begin{center}
\includegraphics[
height=3.2944in,
width=4.9848in
]%
{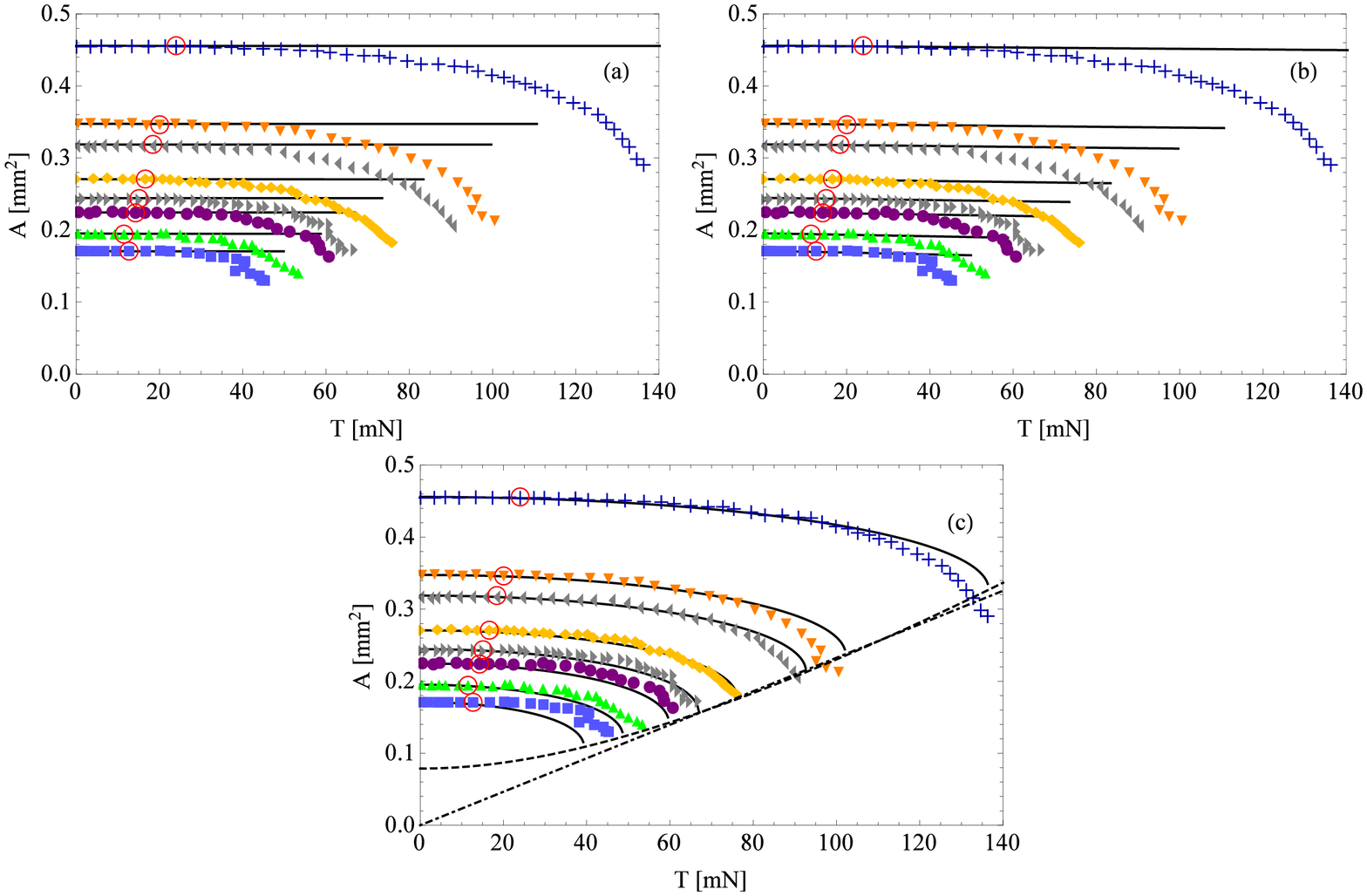}%
\end{center}

Fig. 8 - Contact area as a function of tangential load for the curves with
normal load $\widetilde{P}=\left[
30.4,15.0,11.5,6.2,3.7,2.0,-0.23,-1.79\right]  $ using $\lambda=0.0023$ for
(a) Johnson (1996) model (b)\ Waters-Guduru model and (c) proposed model.
Solid lines are the theoretical predictions while colored symbols stand for
the experimental results (from Mergel \textit{et al. }(2018), their Fig. 5
(c)). In all the panels (a, b, c) the points corresponding to $\psi=\frac
{3\pi}{8}$ are highlighted with a red circle. The same red circle is shown in
Fig. 7. In panel (c) dot-dashed line stands for $T_{full}=\tau_{0}A$
($\tau_{0}=0.43$ MPa, as reported in Mergel \textit{et al. }(2018)) while
dashed line stands for eq. (\ref{Acr}).

\end{center}

We then move to discuss the original data of Waters and Guduru (2010), which
unfortunately were reported only for initial decay of the contact area, where
they observed a strictly axisymmetric configuration. Fitting the mode mixity
function, we find no evidence that our proposed model "c" should be any better
than their model "b". Our best fit gives for the model "c" $\lambda=0.2331$,
while for Waters-Guduru model we confirm the same $\lambda=0.30$ which they
suggest\footnote{Notice that our definition of $\lambda$ is twice the
$\lambda$ in Waters and Guduru (2010) paper.}. Notice that here we only have
data up to $f\left(  \psi\right)  \approx3$ and mode mixity angle of $3/8\pi$,
so the comparison of the models ends much earlier than in the previous
comparison with experimental data (see Fig. 9). It is however in the
comparison of the contact area decay with tangential load where some
(possible) improvement emerges (Fig. 10). Clearly, the data show a much better
overall agreement with the model.%

\begin{center}
\includegraphics[
height=3.3832in,
width=4.9848in
]%
{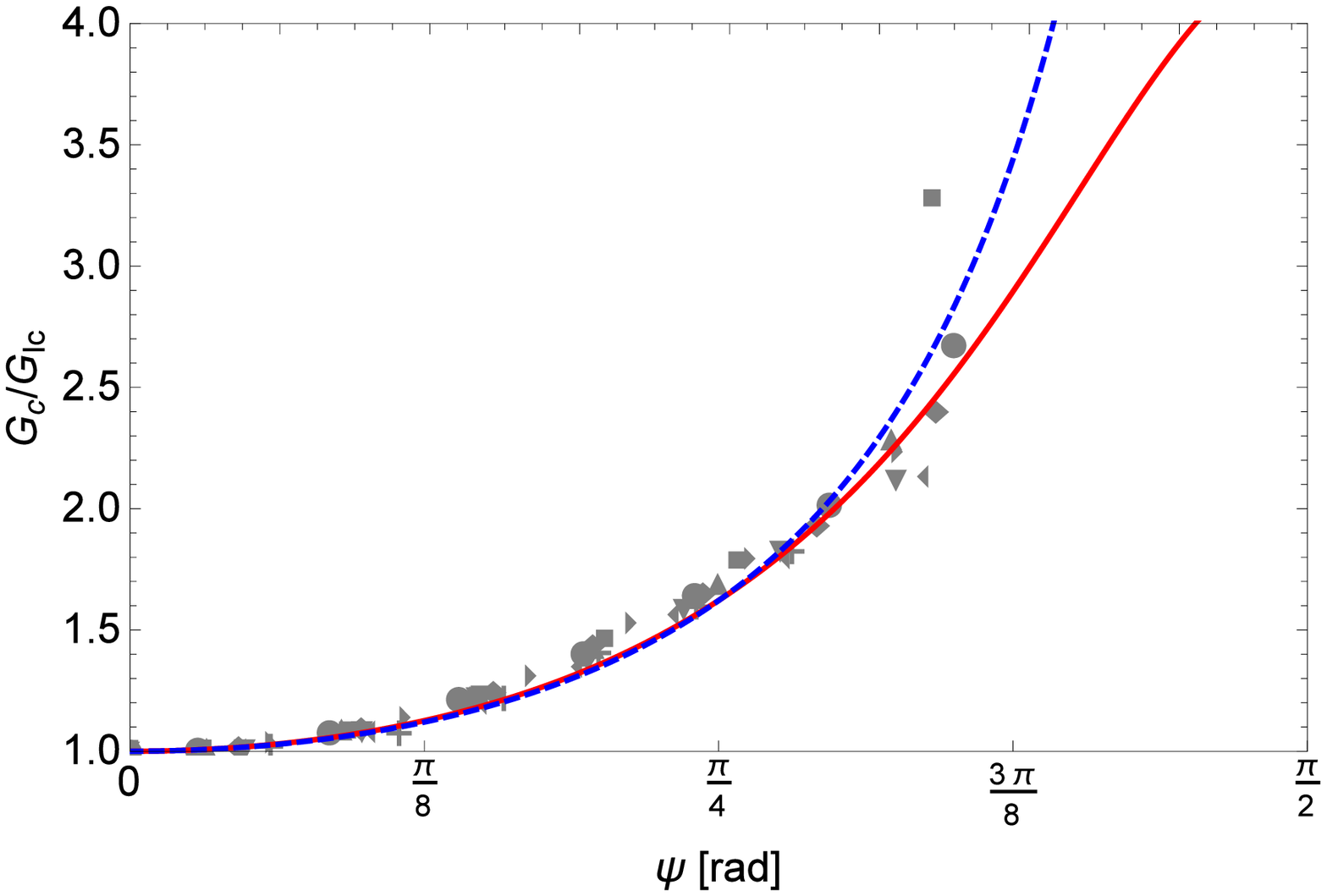}%
\end{center}

\begin{center}
Fig. 9 - $G_{c}/G_{Ic}$ is plotted for the experimental data (gray symbols) in
Waters and Guduru (2010) (data extracted from their Fig. 8). The best fit
procedure provided $\lambda=0.2331$ for the model (c) (solid line). For Waters
and Guduru model "b" (dashed line) we used their fit $\lambda=0.30$ (notice
that our definition of $\lambda$ is twice the $\lambda$ in Waters and Guduru
(2010) paper)$.$%

\begin{center}
\includegraphics[
height=2.5112in,
width=4.9848in
]%
{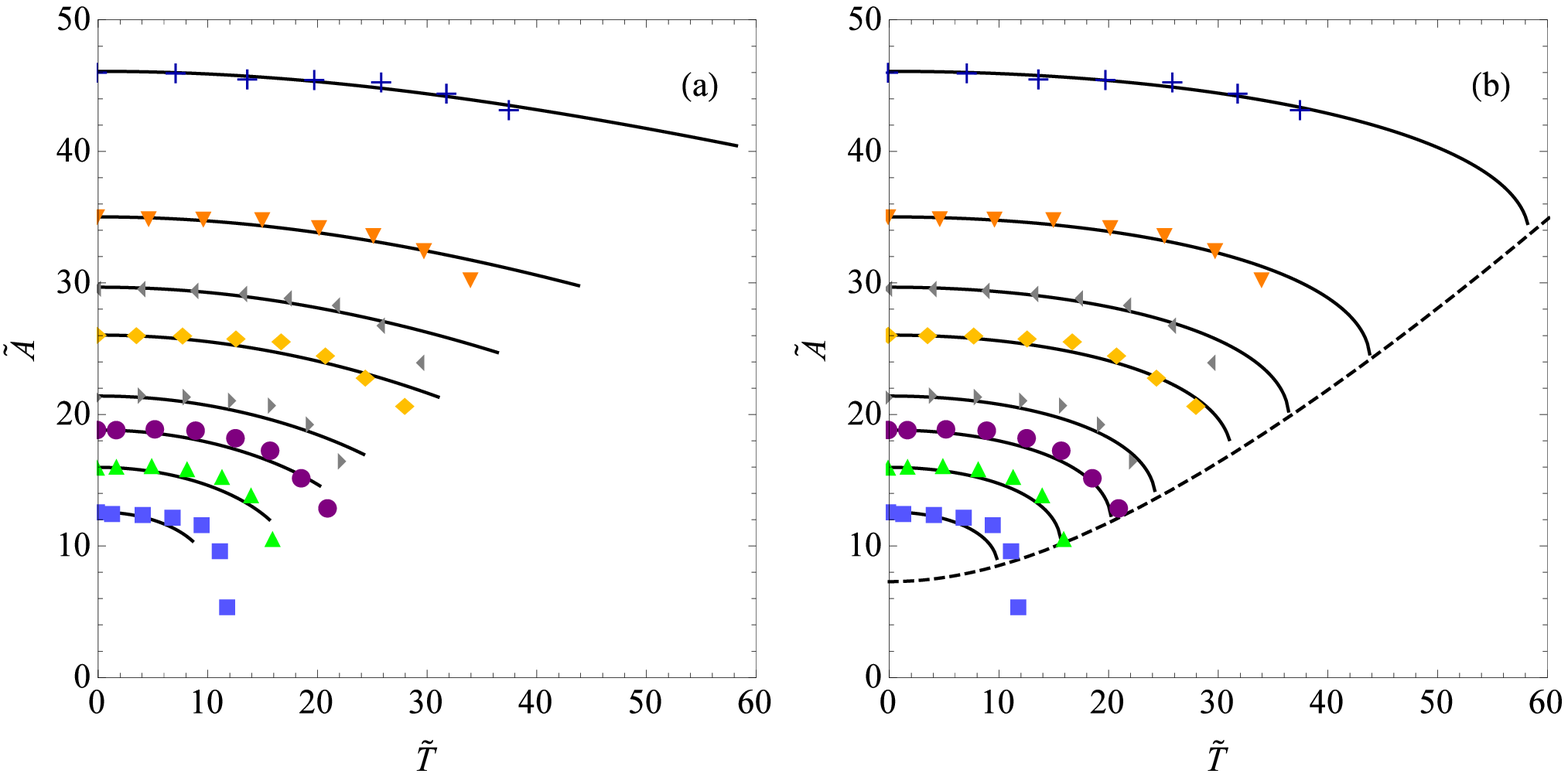}%
\end{center}

Fig. 10 - Contact area as a function of tangential load for Waters and Guduru
(2010) data (their Fig. 8) in dimensionless form. (a) model "b" by Waters and
Guduru, (b) our proposed model "c".
\end{center}

As a way to investigate if the {jump instabilities} we predict in
our model are related to the cycles of detachment/reattachment in
Waters-Guduru (2010), in the absence of better data (the authors report mostly
results at moderate loads because of their choice of plotting only
axisymmetric results), we plot in Fig. 11 the ratio between tangential force
$T$ and the Hertz contact area $A_{H}$ at the given value of normal load,
which we could gather from their Fig. 6 (here reproduced as our Fig. 1). For
the tangential force, we used two set of data: the maximum (squares in Fig. 1)
and minimum (circles in Fig. 1) tangential forces in the first cycle of
detachment/reattachment. {Accordingly with our previous analysis,
the minima of tangential loads should correspond to full-sliding solution with
a Hertzian value of contact area, which, in turns, should approximately
correspond along with a "material constant" shear strength $\tau_{0}$, as it
is often measured for soft materials.} Fig. 11 shows indeed that the value
obtained for the average shear stress is nearly constant, whereas the same
hypothesis does not work for the points at the maxima of tangential load.
Indeed for low normal load the data suggest more an unstable jump rather than
a smooth transition to full sliding in accordance to the proposed model "c".%
\begin{center}
\includegraphics[
height=3.3636in,
width=4.9848in
]%
{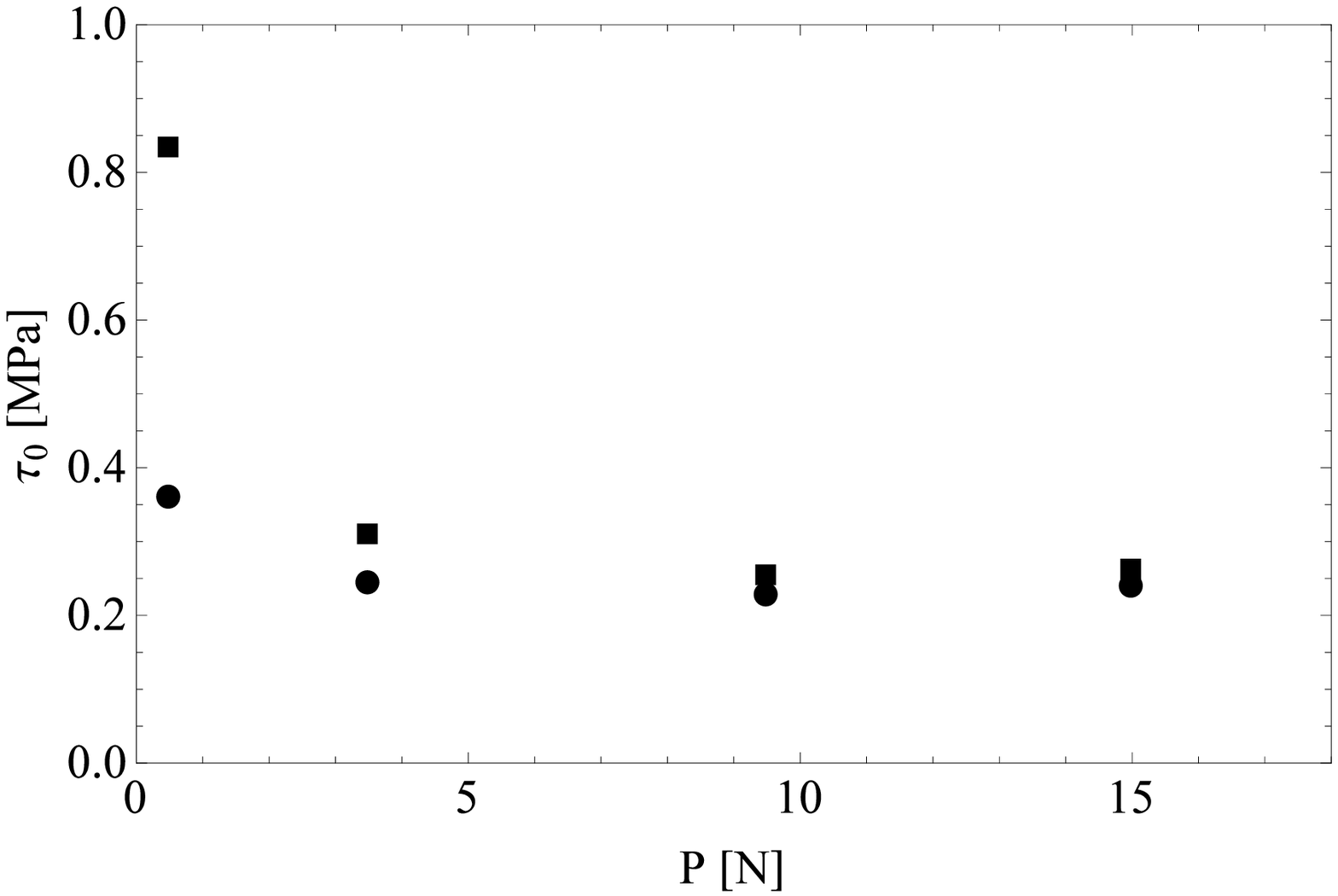}%
\end{center}

\begin{center}
Fig. 11 - Average shear stress computed as tangential force divided by
Hertzian contact area as a function of normal load in Waters and Guduru (2010)
data (data extracted from their Fig. 6, our Fig. 1). Squares refer to the
maximum tangential forces (corresponding to the squares in Fig. 1), circles to
the minimum tangential forces (corresponding to the circles in Fig. 1) in the
first cycle of detachment/reattachment.
\end{center}

\section{Conclusions}

In this work, we have discussed the implications of adopting different
mode-mixity functions to account for the dependence of the interfacial
toughness on the phase angle $\psi=\arctan\left(  \frac{K_{II}}{K_{I}}\right)
$ in an adhesive contact subjected to tangential load. Three different models
have been compared, all originally proposed by Hutchinson \& Suo (1992) in the
framework of fracture mechanics of bi-material interfaces, but of which only
two have been used so far, respectively by Johnson (1996) and Waters \& Guduru
(2010) in the particular problem of contact area shrinking under shear loads.
{We showed that changing the mode-mixity function different contact
behaviors can be obtained, particularly for high phase angles $\psi>3\pi/8$,
values that are quite common in the experiments reported in the literature
(Waters \& Guduru (2010), Mergel \textit{et al.} (2018)). It was shown that,
depending on the interfacial properties, smooth shrinking of contact area up
to full sliding or abrupt drop to }the Hertzian {solution through a jump
instability may be observed which depends on the type of load control. }

Neither Johnson (1966) nor Waters \& Guduru (2010) models fit Mergel
\textit{et al.} (2018) experimental results accurately, because the contact
area remains almost constant for moderate loads. The model we propose fits
reasonably well the data, and the locus of the jump instability points turns
out to be extremely close to the strength criterion condition $\widetilde
{T}_{full}=\widetilde{\tau}_{0}\widetilde{A}$ suggested by Mergel \textit{et
al. }(2018). In the data of Waters \& Guduru (2010), our model seems to lead
more naturally to the jump instability points, and the average shear stress
$\widetilde{\tau}_{0}$ at the minima of tangential load seem to be indeed a
"material constant", as it is usually observed for soft materials. More
investigations are needed to ascertain which conditions are to be expected in
the transition to sliding of soft materials, but it is clear that the
influence of the mixed mode function at high mode mixity is crucial.

\section*{Author Contribution Statement}

AP and MC contributed equally to this work. 

\section*{Acknowledgements}

A.P. is thankful to the DFG (German Research Foundation) for funding the
projects HO 3852/11-1 and PA 3303/1-1.

\section*{References}

Cao, H. C., \& Evans, A. G. (1989). An experimental study of the fracture
resistance of bimaterial interfaces. Mechanics of materials, 7(4), 295-304.

Ciavarella, M. (2018). Fracture mechanics simple calculations to explain small
reduction of the real contact area under shear. Facta universitatis, series:
mechanical engineering, 16(1), 87-91.

Ciavarella, M., \& Papangelo, A. (2017). Discussion of \textquotedblleft
Measuring and Understanding Contact Area at the Nanoscale: A
Review\textquotedblright(Jacobs, TDB, and Ashlie Martini, A., 2017, ASME Appl.
Mech. Rev., 69 (6), p. 060802). Applied Mechanics Reviews, 69(6), 065502.

Homola, A. M., Israelachvili, J. N., McGuiggan, P. M., \& Gee, M. L. (1990).
Fundamental experimental studies in tribology: The transition from
\textquotedblleft interfacial\textquotedblright\ friction of undamaged
molecularly smooth surfaces to \textquotedblleft normal\textquotedblright%
\ friction with wear. Wear, 136(1), 65-83.

Hutchinson, J. W. (1990). Mixed mode fracture mechanics of interfaces.
Metal-ceramic interfaces, 4, 295-306.

Hutchinson, J. W. \& Suo, Z. (1992). Mixed mode cracking in layered materials.
In Advances in applied mechanics, vol. 29 (eds J. W. Hutchinson \& T. Y. Wu),
pp. 63--191. Boston, MA: Academic Press.

Jacobs, T. D. B., \& Martini, A. (2017). Measuring and understanding contact
area at the nanoscale: a review. Applied Mechanics Reviews, 69(6), 060802.

Johnson, K. L., Kendall, K. \& Roberts, A. D. (1971) Surface energy and the
contact of elastic solids. Proc. R. Soc. Lond. A 324, 301--313.

Johnson, K. (1985). Contact Mechanics. Cambridge: Cambridge University Press. doi:10.1017/CBO9781139171731

Johnson, K. L., (1997), Adhesion and friction between a smooth elastic
spherical asperity and a plane surface. In Proceedings of the Royal Society of
London A453, No. 1956, pp. 163-179).

Johnson KL, (1996), Continuum mechanics modeling of adhesion and friction.
Langmuir 12:4510--4513.

Maugis, D. (2013). Contact, adhesion and rupture of elastic solids (Vol. 130). Springer Science \& Business Media.

Mergel, J. C., Sahli, R., Scheibert, J., \& Sauer, R. A. (2018). Continuum
contact models for coupled adhesion and friction. The Journal of Adhesion,
DOI: 10.1080/00218464.2018.1479258

Papangelo, A., Ciavarella, M., \& Barber, J. R. (2015). Fracture mechanics
implications for apparent static friction coefficient in contact problems
involving slip-weakening laws. Proc. R. Soc. A, 471(2180), 20150271.

Pastewka, L., \& Robbins, M. O. (2014). Contact between rough surfaces and a
criterion for macroscopic adhesion. Proceedings of the National Academy of
Sciences, 201320846.

Sahli, R., Pallares, G., Ducottet, C., Ben Ali, I. E., Al Akhrass, S.,
Guibert, M., Scheibert, J. (2018). Evolution of real contact area under shear,
Proceedings of the National Academy of Sciences, 115 (3) 471-476; DOI: 10.1073/pnas.1706434115

Savkoor, A. R. \& Briggs, G. A. D. (1977). The effect of a tangential force on
the contact of elastic solids in adhesion. Proc. R. Soc. Lond. A 356, 103--114.

Schallamach, A. (1971). How does rubber slide?. Wear, 17(4), 301-312.

Vakis, A. I., Yastrebov, V. A., Scheibert, J., Minfray, C., Nicola, L., Dini,
D., ... \& Molinari, J. F. (2018). Modeling and simulation in tribology across
scales: An overview. Tribology International 125, 169-199. DOI: 10.1016/j.triboint.2018.02.005

Waters JF, Guduru PR, (2010), Mode-mixity-dependent adhesive contact of a
sphere on a plane surface. Proc R Soc A 466:1303--1325.

Yoshizawa, H., Chen, Y. L., \& Israelachvili, J. (1993). Fundamental
mechanisms of interfacial friction. 1. Relation between adhesion and friction.
The Journal of Physical Chemistry, 97(16), 4128-4140.

\end{document}